\documentclass[sigconf]{acmart}

\usepackage{booktabs} 

\setcopyright{rightsretained}

\acmDOI{10.475/123_4}

\acmISBN{123-4567-24-567/08/06}

\acmConference[WSDM'19]{The 12th ACM International Conference on Web Search and Data Mining}{February, 2019}{Australia}
\acmYear{2018}
\copyrightyear{2018}

\acmArticle{4}
\acmPrice{15.00}

\begin{document}
\title{Modelling Sequential Music Track Skips using a Multi-RNN Approach}

\author{Christian Hansen}
\affiliation{%
  \institution{University of Copenhagen}
  \department{Department of Computer Science}
}
\email{chrh@di.ku.dk}

\author{Casper Hansen}
\affiliation{%
  \institution{University of Copenhagen}
  \department{Department of Computer Science}
}
\email{c.hansen@di.ku.dk}

\author{Stephen Alstrup}
\affiliation{%
  \institution{University of Copenhagen}
  \department{Department of Computer Science}
}
\email{s.alstrup@di.ku.dk}

\author{Jakob Grue Simonsen}
\affiliation{%
  \institution{University of Copenhagen}
  \department{Department of Computer Science}
}
\email{simonsen@di.ku.dk}

\author{Christina Lioma}
\affiliation{%
  \institution{University of Copenhagen}
  \department{Department of Computer Science}
}
\email{c.lioma@di.ku.dk}

\begin{abstract}
Modelling sequential music skips provides streaming companies the ability to better understand the needs of the user base, resulting in a better user experience by reducing the need to manually skip certain music tracks. This paper describes the solution of the University of Copenhagen ''DIKU-IR'' team in the ''Spotify Sequential Skip Prediction Challenge'', where the task was to predict the skip behaviour of the second half in a music listening session conditioned on the first half. We model this task using a Multi-RNN approach consisting of two distinct stacked recurrent neural networks, where one network focuses on encoding the first half of the session and the other network focuses on utilizing the encoding to make sequential skip predictions. The encoder network is initialized by a learned session-wide music encoding, and both of them utilize a learned track embedding. Our final model consists of a majority voted ensemble of individually trained models, and ranked 2nd out of 45 participating teams in the competition with a mean average accuracy of 0.641 and an accuracy on the first skip prediction of 0.807. Our code is released at \url{https://github.com/Varyn/WSDM-challenge-2019-spotify}.
\end{abstract}


\maketitle

\section{Introduction}
A challenge for content providers is to model how a given user will react to some content, such that users are provided with content that elicits some positive reaction. Spotify, a music streaming company, have the problem of incorporating the sequential nature of a listening session to predict whether a user will skip a given music track or not\footnote{\url{ https://www.crowdai.org/challenges/spotify-sequential-skip-prediction-challenge}}. To this end, a set of approximately 130 million labelled user sessions has been released, where each session consists of 10 to 20 playback tracks. The task is to predict which music tracks a user will skip in the second half of session conditioned on the first half.
This problem can be considered a type of session based recommendation, since no explicit user profile is available, and the user preferences should therefore be estimated within the first half of the session. To capture the dynamics of the session, models based on recurrent neural networks (RNNs) have seen much popularity for session based recommender systems, as they are able to encode temporal information well \cite{hidasi2015session,tan2016improved, quadrana2017personalizing}.

In this paper we present our solution to the "Spotify Sequential Skip Prediction Challenge", which ranked second among all 45 submitted solutions. Our solution is based on using two distinct stacked RNNs, one stacked RNN to encode the first half of a session, and one stacked RNN responsible for making the skip predictions conditioned on the state of the first RNN. The stacked RNN used for encoding the first half of the session is initialized using meta features related to the session as a whole, and an encoding of all tracks listened to in the entire session. Both RNNs utilize a track embedding based on provided track features and a learned embedding. Our model shares similarities with architectures used in sequence-to-sequence networks \cite{sutskever2014sequence}, which also consist of two distinct RNNs for encoding and decoding. Sequence-to-sequence networks are especially popular for machine translation where a strong encoding of the sentence is needed \cite{wu2016google, gehring2017convolutional}.

\section{Sequential Music Skip Prediction}
In this section we present our method. First we present the Spotify dataset \cite{brost2019music} and its features for both the sessions and the tracks. We thereafter present an overview of the model, and then go into greater detail of each component of the model.  

\subsection{Features}
\label{s:Features}
We first describe the dataset and features to establish a common terminology to use throughout the paper, and then the feature engineering used to process the data. The dataset consists of two primary parts consisting of a user log and a track dataset.

\textbf{The user log} contains the user sessions, which for each user is a sequence of track playbacks. 
A track playback contains information about the user in general (e.g., if they are premium or the day of the week they are listening), features related to a single track playback (e.g., what action the user took to end up listening to the current track, or which action the user took to stop listening to this track), and lastly the id of the track being listened to.
The training data contains the playback track features for all tracks in the user session. For the testing data this is available for the first half of the session, while the last half of the session only contains a track id and position in the session of the track being played.

\textbf{The track dataset} contains a number of features for each track which both relate to meta information about the song (e.g., popularity and release information) and specific information related to the musical content of the song (e.g., beat strength or flatness). 

\subsubsection{Feature processing}
Our method is not reliant on any extensive data pre-processing, so the data preparation is straight forward: the user session is represented as 3 types of data:
\begin{enumerate}
    \item Meta information associated with the whole session
    \item A sequence of playback tracks for the first half of the listening session
    \item A sequence of track ids and position in the overall session for the second half of the listening session. This was done to mimic how the testing data was constructed.
\end{enumerate}
The meta information (1) for the whole session consists of whether the user is a premium user, the length of the session, and the day of the week. These are encoded separately using a one hot encoding. 
The first half of the listening session (2) contains all the features for each playback track, except for the features listed in the meta information (1). All categorical features are encoded using a one hot encoding.
The second half of the listening session (3) only contains the track id and position in the session for each track.

The track data is standardized such that each feature has 0 mean and unit variance, and is otherwise used as is for representing a track. 

\begin{figure*}
    \centering
    \vspace{-7pt}
    \includegraphics[width=0.9\linewidth]{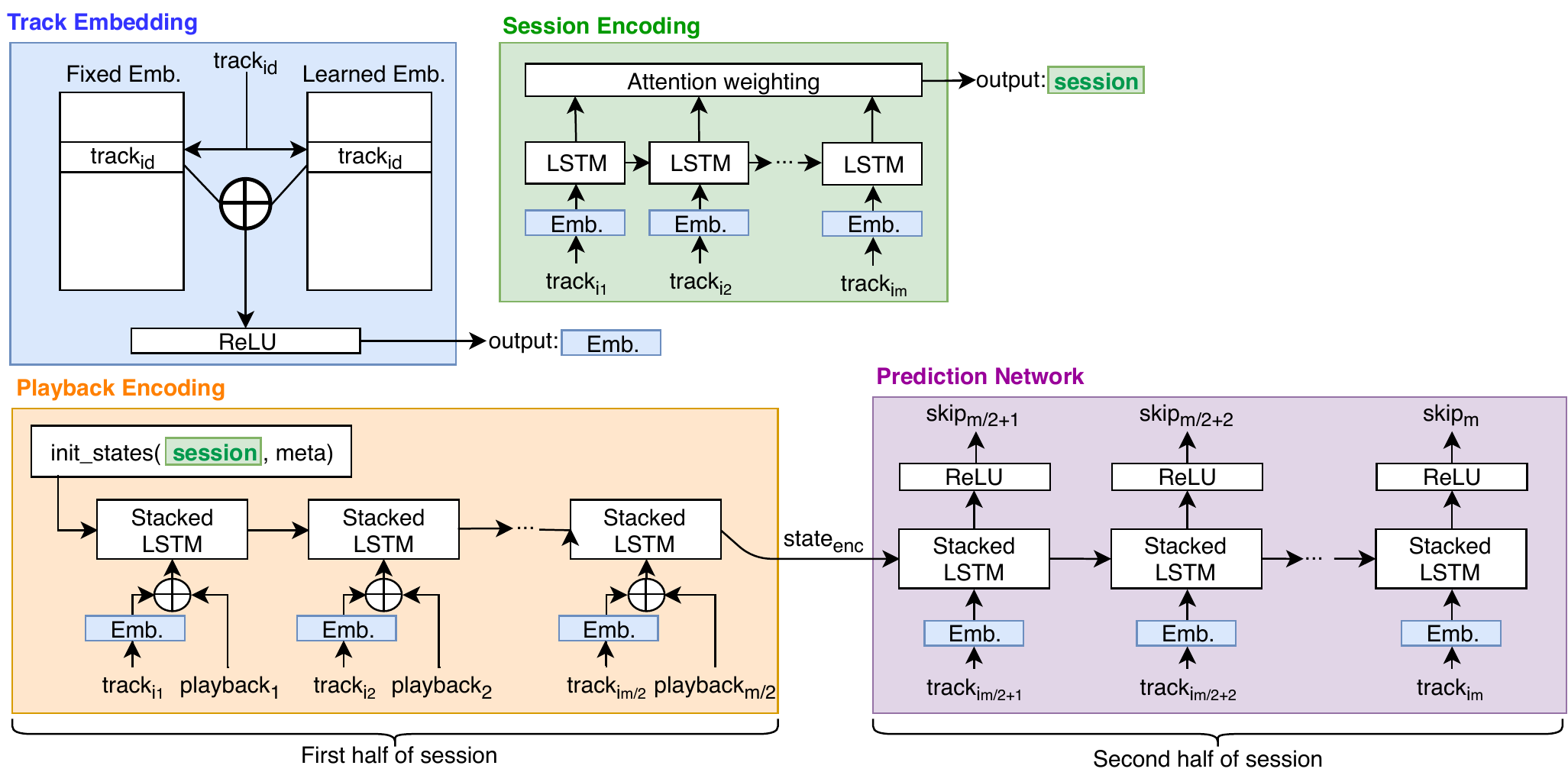}
    \caption{Network architecture. $\oplus$ represents vector concatenation of the two embedding look-ups in the Track Embedding and between an embedded track and playback information in the Playback Encoder.}
    \vspace{-12pt}
    \label{fig:model}
\end{figure*}

\subsection{Model}
This section will explain the neural architecture of our model, and the network can be seen in Figure \ref{fig:model}. Overall, the network consists of 4 parts: 1) An embedding of the tracks; 2) A network for encoding all tracks in the full session; 3) A network for encoding the playback track sequence, which is the first half of a session; and lastly 4) A prediction network that takes the encoding from (3) and makes a skip prediction for each track of the second half of the session.

\subsubsection{Track embedding}
The purpose of the track embedding is to produce an embedding of the track that both utilizes the provided features, but also allows the network to learn an embedding specifically optimized for the skip prediction task. These two kinds of embeddings are concatenated into a single embedding.

The track embedding has 2 tunable parameters: the size of the learned embedding for the track and the size of the final track embedding. Each track has an identifier, \(i\), which is used to index into two embedding matrices \(E_{fixed}\) and \(E_{learned}\). \(E_{fixed}\) contains all features from the track data, which have been normalized as described in Section \ref{s:Features}. \(E_{learned}\) is a learned embedding, which is initialized using a uniform distribution with values from $[-0.05, 0.05]$. The final track embedding is then computed as:
\begin{align}
        \textbf{\textrm{track}}_i &= \textrm{ReLU} \big( W_t [E_{\textrm{fixed}} \textbf{w}_i \oplus E_{\textrm{learned}} \textbf{w}_i  ] + \textbf{b}_t \big)
\end{align}
where \textrm{ReLU} is the Rectified Linear Unit activation function, \(\oplus\)  corresponds to vector concatenation, and \(\textbf{w}_i\) is the identity vector whose \(i\)th entry is $1$ and all other entries 0. The size of \(\textbf{\textrm{track}}_i\) is the final track embedding size.

\subsubsection{Session encoding}
The purpose of making an initial encoding of the whole session (consisting of all tracks), is to allow the network to be able to learn what kind of music is listened to across the session. The goal of this is to allow the network to utilize variation, genre, and general music similarity throughout the session.
To do this, the track embedding presented in the previous section is used. For a session we have a sequence of embedded tracks \([\textbf{\textrm{track}}_{i_1},\textbf{\textrm{track}}_{i_2},...,\textbf{\textrm{track}}_{i_m}]\), where $i_1$ is the track embedding of the first track of the session and \(m\) is the length of the session. We first apply a RNN with Long Short-Term Memory (LSTM) units \cite{hochreiter1997long}, and get the output produced for each timestep:
\begin{align}
    [\textbf{o}_{1},\textbf{o}_{2},...,\textbf{o}_{m}] = \textrm{LSTM}([\textbf{\textrm{track}}_{i_1},\textbf{\textrm{track}}_{i_2},...,\textbf{\textrm{track}}_{i_m}])
\end{align}
where \(\textbf{o}_i\) is the \(i\)th output of the LSTM, and \(\textrm{LSTM}(\cdot)\) is the application of the LSTM on the whole sequence. The final session embedding is made using an attention-weighted sum over the outputs,
\begin{align}
    \textbf{\textrm{session}} = \sum_{i=1}^m  \textbf{o}_i \frac{\textrm{exp}(W_a \textbf{o}_i + b_a)}{\sum_{j=1}^m \textrm{exp}(W_a \textbf{o}_j + b_a)}
\end{align}
where the second term of the sum corresponds to the attention weighting of the $i$th output, \(\textrm{exp}\) is the exponential function, and \(\textbf{\textrm{session}}\) is the session embedding. The size of the session embedding is dependent on the size of the LSTM unit.

\subsubsection{Playback encoding}
The purpose of the playback encoding network is to make an encoding of the first half of the session, and then use this encoding when making the skip prediction for each track in the second half of the session.

The encoding network uses a stacked RNN with a depth of 2 using LSTM units. The initial state of the LSTM is given by 4 linear fully connected layers taking as input the encoding of all tracks and the session meta information, and producing the initial hidden state and output of the LSTM, which we refer to as hidden and output respectively:
\begin{align}
    \textrm{\textbf{hidden}}_l^{\textrm{initial}} &= W_{s,l} [\textbf{m}  \oplus \textbf{\textrm{session}} ] + \textbf{b}_{s,l} \\
    \textrm{\textbf{output}}_l^{\textrm{initial}} &= W_{o,l} [\textbf{m}  \oplus \textbf{\textrm{session}} ] + \textbf{b}_{o,l}
\end{align}
where \(l \in \{1,2\}\) indicates the first or second LSTM layer. For later use we will simply refer to the initial state of the whole stacked LSTM as \(\textrm{state}^{\textrm{initial}}\). \(\textbf{m}\) is the meta information associated with the session as described in 
Section \ref{s:Features}. The input to the stacked LSTM at the \(j\)th timestep for the first half of the session, $\textbf{s}^{\textrm{f}}_j$, is constructed as:
\begin{align}
    \textbf{s}^{\textrm{f}}_j = [\textbf{\textrm{track}}_{i_j} \oplus \textbf{\textrm{playback}}_j]
\end{align}
where \(\textbf{\textrm{playback}}_j\) is the playback track features described in Section \ref{s:Features}, for the \(j\)th track in the session. The sequence for the first half is then \([\textbf{s}^{\textrm{f}}_1,\textbf{s}^{\textrm{f}}_2,...,\textbf{s}^{\textrm{f}}_{\frac{m}{2}}]\). The playback encoding is the final state of the LSTM after the whole sequence has been read:
\begin{align}
    \textrm{state}_{\textrm{enc}} = \textrm{STACK-LSTM}_{\textrm{enc}}([\textbf{s}^{\textrm{f}}_1,\textbf{s}^{\textrm{f}}_2,...,\textbf{s}^{\textrm{f}}_{\frac{m}{2}}]|\textrm{state}^{\textrm{initial}})
\end{align}
where \(\textrm{state}_{\textrm{enc}}\) is the final state of the stacked LSTM, and STACK-\(\textrm{LSTM}_{\textrm{enc}}(\cdot)\) is the application of the stacked LSTM on the input sequence conditioned on the initial state, \(\textrm{state}^{\textrm{initial}}\). The size of the LSTMs in both layers are chosen to be the same.

\subsubsection{Prediction network}
The purpose of the prediction network is to make a prediction for each track in the second half of the session. This is done by using a stacked LSTM that reads the second half of the session and produces an output for each track, which is used to make the final predictions. Note that the stacked LSTMs in the encoder and the prediction network have the same LSTM size, but do not share any weights. The input at timestep \(j\) for the stacked LSTM in the prediction network is a track embedding concatenated with the position of the track in the full session (\(j+ \frac{m}{2}\)). We denote the input sequence to the stacked LSTM as: \([\textbf{s}^s_1,\textbf{s}^s_{2},...,\textbf{s}^s_{\frac{m}{2}}]\) and compute the stacked LSTM as,
\begin{align}
[\textbf{o}_{1}^s,\textbf{o}_{2}^s,...,\textbf{o}_{\frac{m}{2}}^s] = \textrm{STACK-LSTM}_{\textrm{pred}}([\textbf{s}^s_1,\textbf{s}^s_{2},...,\textbf{s}^s_{\frac{m}{2}}]|\textrm{state}_{\textrm{enc}})    
\end{align}
The skip prediction for each track is then given as the output of two fully connected layers, where the first layer has the same size as the LSTM and a ReLU activation function, whereas the second layer produces the prediction via a sigmoid activation function. 

The whole network is trained using binary cross entropy, with the ground truth skip behavior of the user as the target for each prediction.

\section{Experimental evaluation} \label{s:experimental-evaluation}
We now describe the experimental evaluation of our model. We first describe the performance metric and simple baselines provided by the competition organizers, followed by how our model was tuned as well as implementation details. We then study different configurations of our model on a validation set, and lastly report the test performance of our final model.

\subsection{Performance metric} \label{ss:performance-metric}
The competition employs the Mean Average Accuracy as the official performance metric. The average accuracy is computed for each session and then averaged across all sessions. The average accuracy is defined as:
\begin{equation}
    AA = \frac{\sum_{i=1}^T A(i)L(i)}{T}
\end{equation}
where $T$ is the number of tracks to be predicted in a session, $A(i)$ is the accuracy at position $i$ of the track sequence, and $L(i) = 1$ if the prediction at position $i$ was correct,
and $L(i) = 0$ otherwise. Additionally, for the test performance we also report the accuracy on the first skip prediction, as that was used for breaking potential ties among the participants.

\subsection{Baselines}\label{ss:baselines}
For comparison we include the 3 provided baselines by the competition organizers; the baselines are relatively simple, but show the performance using intuitive rules for making the skip decisions:
\begin{enumerate}
    \item Predict all tracks to be skipped;
    \item Predict track to be skipped if its skip rate in the training set was greater than 0.5;
    \item Predict the last action from first the half of the session for all tracks in the second half of the session.
\end{enumerate}
We report the performance metrics on these baselines for the final comparison on the test set and refer to them as Baseline 1-3.

\subsection{Tuning} \label{ss:tuning}
We measure the effectiveness of our model using mean average accuracy and maximize this metric for tuning parameters. For training we randomly shuffle the provided dataset of 130 million sessions, and set aside 0.2 million sessions for validation, which will be used to detect overfitting. Testing is done on a separate dataset consisting of 31.3 million sessions.

Due to the massive size of the dataset we fix the network parameters with a track embedding of size 50; a fully connected layer with size 350 for combining the fixed track embedding with the learned track embedding; a LSTM size of 100 for the encoding of tracks in a session; an LSTM size of 500 in both the encoder and predictor networks; and lastly a fully connected layer with size 500 for making the skip prediction of the predictor network output. These were initially decided based on rapid tests trained for a short number of iterations, but due to computational limitations we did not explore them in full detail. For the network training parameters we explored batch sizes in the set $\{50, 100, 200, 300, 400\}$, and used a learning rate of 0.0005. For training the network we used the Adam optimizer \cite{kinga2015method} with a learning rate of 0.0005. We trained the models using Titan X GPUs, and used 40-60 hours to fully train each model.

\subsection{Implementation details}
Due to the massive size of the dataset we implemented the LSTMs in our model using \texttt{cuDNNLSTM} in TensorFlow \cite{abadi2016tensorflow}, which is up to 7.2 times faster than traditional LSTM implementations \cite{lstmSpeed}. However, this implementation requires a fixed amount of time steps, which was set to 10 for both RNNs. To handle sessions of varying length we applied pre-padding on the input to the playback encoder network and post-padding to the input to the prediction network, such that all inputs fitted the fixed size.

\subsection{Results} \label{ss:results}
Table \ref{tab:batchsize_val} displays the validation mean average accuracy of our model when varying the batch size, which we found to be one of the most influential parameters to tune for our network architecture. The table shows that a larger batch size leads to a higher average accuracy with a batch size of 300 performing the best.

Table \ref{tab:test} shows the test mean average accuracy and accuracy of the first skip prediction. The table shows the performance of the three provided baselines, our best performing submission of a single model (using a batch size of 300), as well as a majority voting among the 5 models with varying batch size. We observed that the average correlation between the predictions made by the 5 models was 0.851, and created a final model by taking a majority voting among the models. The majority voted model performed better than the best single model with an average accuracy of 0.641 and first skip prediction accuracy of 0.807, which was also notably better than the three baselines. On the final leaderboard this ranked as the second best submission in the competition.

\begin{table}
    \centering
    \scalebox{0.9}{
    \begin{tabular}{lc}
    \toprule
         Batch size & Validation AA \\ \hline
         400 & 0.637 \\ 
         300 & \textbf{0.638} \\
         200 & 0.635 \\
         100 & 0.633 \\
         50 & 0.631 \\
        \bottomrule
    \end{tabular}}
    \caption{Model validation performance}
    \label{tab:batchsize_val}
    \vspace{-22pt}
\end{table}

\begin{table}
    \centering
    \scalebox{0.9}{
    \begin{tabular}{lcc}
    \toprule
         Model & Test AA & First Prediction Accuracy \\ \hline
         Baseline 1 & 0.405 & 0.541 \\
         Baseline 2 & 0.409 & 0.559 \\
         Baseline 3 & 0.537 & 0.742 \\ \hline
         Our model & 0.638 & 0.805 \\ 
         Our majority voted model & \textbf{0.641} & \textbf{0.807} \\
        \bottomrule
    \end{tabular}}
    \caption{Model test performance
    }
    \label{tab:test}
    \vspace{-22pt}
\end{table}

\section{Conclusion} \label{s:conclusion}
This paper described the participation of the University of Copenhagen ''DIKU-IR'' team in the ''Spotify Sequential Skip Prediction Challenge'', where the task was to predict the skip behaviour of the second half in a music listening session conditioned on the first half. We proposed a new model for the task of modelling sequential music skip behaviour in a given user session of streamed music content. Our model consisted of a Multi-RNN approach with two distinct stacked recurrent neural networks, which can be considered as an encoding and predictor network. The encoder network exploited a learned session-wide encoding of the musical content, while both of them utilized a learned embedding of each musical track. We combined individually trained models with different batch sizes in a majority voted ensemble to obtain a mean average accuracy of 0.641 and an accuracy on the first skip prediction of 0.807, which was the second best submission of the competition among 45 participating teams.

\begin{acks}
Funded by the Innovation Fund Denmark, DABAI project.
\end{acks}

\bibliographystyle{ACM-Reference-Format}
\bibliography{main}

\end{document}